\documentclass[%
 preprint,
 amsmath,amssymb,
 aps,
]{revtex4-1}

\usepackage{graphicx}
\usepackage{hyperref}
\usepackage{float}
\usepackage{dcolumn}
\usepackage{bm}
\usepackage{times}
\usepackage{gensymb}
\usepackage{natbib}
\usepackage{indentfirst}
\usepackage{lineno}
\newcommand{\specialcell}[2][c]{%
  \begin{tabular}[#1]{@{}c@{}}#2\end{tabular}}

\begin{document}

\preprint{}

\title{Glassy dynamics of landscape evolution}

\author{Behrooz Ferdowsi}

\affiliation{%
Department of Geosciences, Princeton University, \\Princeton, NJ 08544, USA. \& \\
 Earth and Environmental Science, University of Pennsylvania, \\Philadelphia, PA 19104, USA \& \\
 National Center for Earth-surface Dynamics 2, University of Minnesota, Third Avenue SE, Minneapolis, MN 55414, USA
}%
\author{Carlos P. Ortiz}%

\affiliation{%
 Earth and Environmental Science and Department of Physics and Astronomy, University of Pennsylvania,\\ Philadelphia,
PA 19104, USA}%

\author{Douglas J. Jerolmack}%
\altaffiliation[corresponding author]{}
\affiliation{%
Earth and Environmental Science, University of Pennsylvania,\\ Philadelphia,
PA 19104, USA}%
 \email{sediment@sas.upenn.edu}

\date{\today}

\begin{abstract} 
\noindent  Soil creeps imperceptibly downhill, but also fails catastrophically to create landslides. Despite the importance of these processes as hazards and in sculpting landscapes, there is no agreed upon model that captures the full range of behavior. Here we examine the granular origins of hillslope soil transport by Discrete Element Method simulations, and re-analysis of measurements in natural landscapes. We find creep for slopes below a critical gradient, where average particle velocity (sediment flux) increases exponentially with friction coefficient (gradient). At critical there is a continuous transition to a dense-granular flow rheology. Slow earthflows and landslides thus exhibit glassy dynamics characteristic of a wide range of disordered materials; they are described by a two-phase flux equation that emerges from grain-scale friction alone. This glassy model reproduces topographic profiles of natural hillslopes, showing its promise for predicting hillslope evolution over geologic timescales.
\end{abstract}
\maketitle 

\noindent  Creep in soil-mantled hillslopes takes place due to slow rearrangements of grains that can be reasonably approximated as a viscous flow \citep{culling1963soil}. At critical slopes or under certain perturbations like rain events, however, soil may fail catastrophically to create landslides \citep{iverson1997physics}. These processes control the erosion and form of hillslopes, and the delivery of sediment to rivers \citep{roering2007functional}. Despite widely varying materials and environments, hillslope soil motion falls into two distinct categories: slowly creeping ``earthflows'' associated with surface velocities that cover ten orders of magnitude up to $\sim 10^{-1}$ m/s, and rapid landslides (including debris flows, mudflows, etc.) that are faster than  $\sim 10^{0}$ m/s \citep{cruden1996landslides,hungr2001review,hilley2004dynamics,saunders1983rates} (Fig. \ref{fig1}C). Much progress has been made in mechanistic models for the latter; in particular, continuum models based on mass and momentum conservation for the granular and fluid phases are able to reproduce important aspects of soil failure and mass-movement runout \citep{iverson1997physics,mangeney2010erosion}. Models for hillslope soil creep, however, lack a mechanistic underpinning. For over 50 years, a heuristic ``diffusive-like law'' --- in which sediment flux $q_s$ $[L^2/T]$ is proportional to topographic gradient $S = \delta z / \delta x$ --- has been employed to model landscape erosion \citep{culling1965theory,roering1999evidence,dietrich2006search}. In order for soil to creep at sub-critical gradients, it is supposed that dilation occurs as a result of (bio-)physical perturbations such as freeze/thaw/swell, rainsplash, tree throw and burrowing animals \citep{heimsath2002creeping,furbish2007rain,gabet2000gopher}. 

Hillslope soil creep has not been connected to the creep phenomenon observed in diverse amorphous and granular systems with a wide range of materials and particle shapes, including dry \citep{komatsu2001creep} and fluid-driven \citep{houssais2015onset,PhysRevE.94.062609} granular flows. These materials belong to a large class of systems --- including glasses, pastes, foams, gels and suspensions --- that are known to exhibit interesting rheological properties termed ``glassy dynamics'' that include: slow dynamics such as compaction \citep{richard2005slow}; hysteresis and history dependence of the static configurations \citep{berg2002glassy}; and intermittency and spatially heterogeneous dynamics \citep{dauchot2011dynamical}. These behaviors are thought to be a natural consequence of two properties shared by all glassy materials: structural disorder and metastability \citep{sollich1997rheology}. In such materials, thermal motion alone is not enough to achieve complete structural relaxation, and consequently relaxation times are extremely large compared to the time scale of a typical experiment. As a result and for practical purposes, glassy materials are non-equilibrium systems with long memory \citep{berthier2003course}. For the evolution of soil-mantled landscapes over geologic time, such glassy dynamics should be relevant.

Dense granular flows on inclined planes are a reasonable idealized model for hillslope soil transport, and the rheology of such flows have been thoroughly examined in simulations and experiments \citep{staron2008correlated,silbert2003granular,louge2003model, lemieux2000avalanches,daerr2001dynamical,pouliquen2002friction,louge2001dense,jop2005crucial, richard2008rheology}. Local \citep{jop2006constitutive,midi2004dense} and then non-local \citep{pouliquen2009non,bouzid2015non,henann2013predictive} constitutive relations developed for dense, steady flows establish that the effective friction is a nonlinear function of the (nondimensionalized) shear rate. We note, however, that most of these studies assume that granular piles are jammed below the angle of repose; i.e., they do not examine sub-critical creep dynamics (for an exception see \citep{zhang2017microscopic}). 
Creep involves grain-scale rearrangement --- due to structural and mechanical disorder in the pack \citep{darnige2011creep,richard2008rheology} --- that induces exponentially small but finite particle velocities below threshold \citep{chauve2000creep}. Although creep has been recognized for some time \citep{komatsu2001creep}, only recently have researchers begun probing the nature of the creep to fluid transition in granular heap flows. 

Experiments performed with an inclined granular layer show localized and isolated events --- microfailures --- in the bulk at inclinations below the bulk angle of repose. As the inclination increases, microfailures occur more frequently until they coalesce to form an avalanche \citep{amon2013experimental}. It has been suggested that this change from creeping to flowing states is a dynamical phase transition \citep{holmes2003dynamic}, that is phenomenologically similar to a liquid-glass transition \citep{reichhardt2016depinning,dauchot2005dynamical,amon2013experimental}. In the athermal granular system, the transition occurs in the vicinity of a critical force (instead of the critical temperature for thermal glasses). One way to probe this transition is to define an order parameter, that describes a dynamical quantity that changes dramatically across the phase transition. Recent simulations have considered the particular case of the onset of erosion of a single layer of monodisperse grains on a substrate; Yan et al. \citep{yan2016model} and Aussillous et al. \citep{aussillous2016scale} showed that this can be mapped to a plastic depinning transition, a generic phase transition associated with irreversible strain and disorder \citep{reichhardt2016depinning}. In their work, sediment flux (current) works as an order parameter that goes sharply to zero at a critical driving force below which flow does not occur \citep{yan2016model, aussillous2016scale}. In the presence of additional noise, however --- which may take the form of internal disorder, sidewalls/boundary effects, or external perturbations --- localized flow may occur below the critical force in amorphous systems \citep{pouliquen2009non,bouzid2013nonlocal,kamrin2015nonlocal,DeGiuli15082017}. Experiments have demonstrated that the onset of erosion in 3-dimensional (3D) granular flows is indeed continuous, accompanied by subcritical flow and creep \citep{houssais2015onset} and exhibits dynamics qualitatively consistent with a liquid-glass transition \citep{amon2013experimental}; however, the dynamical nature of this transition has not yet been examined in such systems. In this paper we conduct 3D numerical experiments of granular heap flows to first document sub-threshold creep, and then demonstrate that the creep to landslide transition is continuous and is quantitatively consistent with a plastic depinning transition. We then present evidence of such glassy dynamics in the soil transport rates and topographic profiles of real hillslopes in nature.

\section*{\hspace{-0.2in}Granular hillslope model}
\begin{figure}[ht]
\centerline{\includegraphics[width=1.0\textwidth]{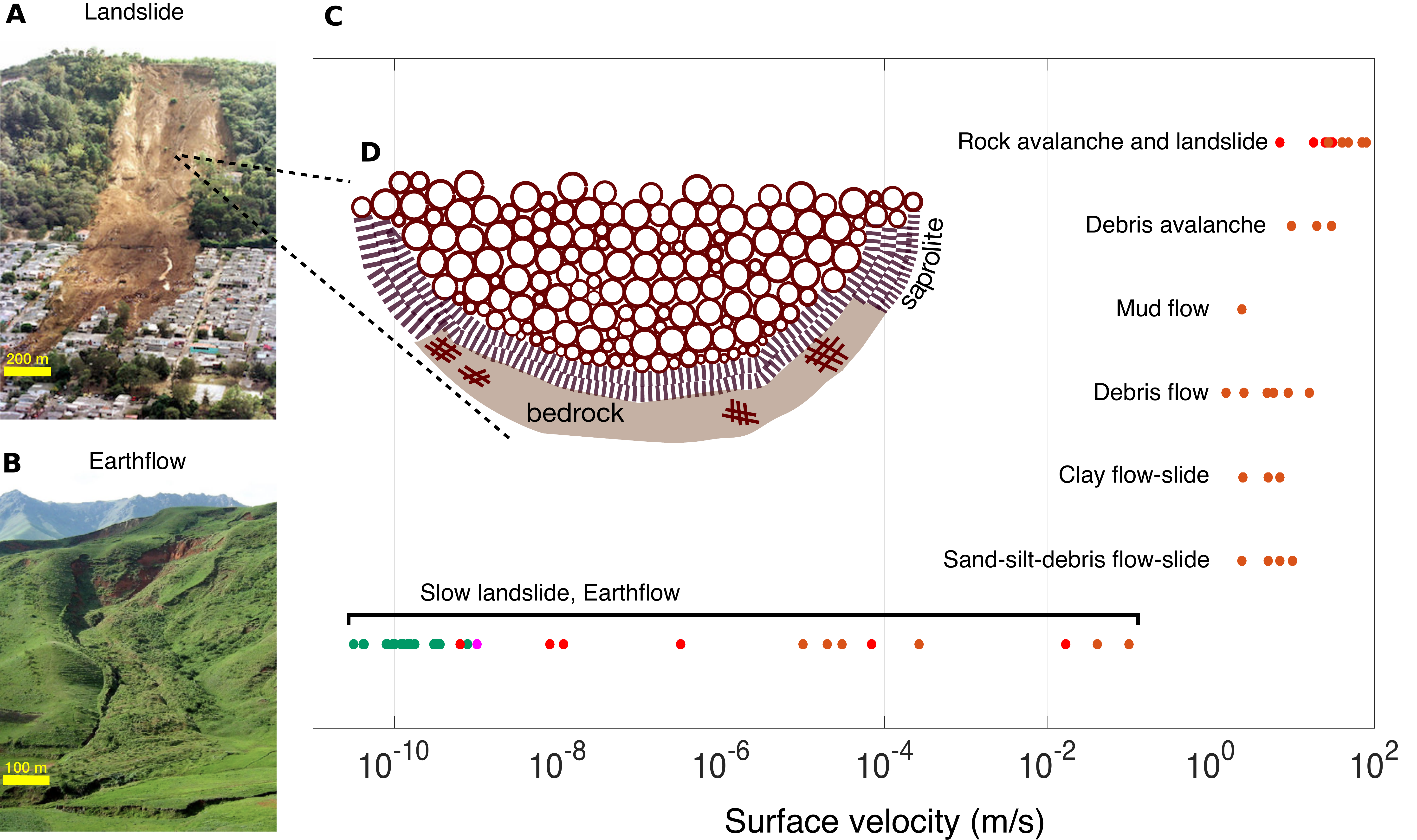}}
\caption{Landslide and creep phenomenology. (A) Rapid landslide in San Salvador, El Salvador; and (B) Slow earthflow in Osh, Kyrgyzstan. (C) Ranges of surface velocities observed for various types of slow and rapid landslides. The datapoints in red, brown, magenta, and green correspond to the observations reported or documented by Cruden \& Varnes (1996)\citep{cruden1996landslides}, Hungr et al. (2001)\citep{hungr2001review}, Hilley et al, (2004)\citep{hilley2004dynamics}, and Saunders \& Young (1983)\citep{saunders1983rates}, respectively. (D) Schematic cross section of a soil-mantled hillslope. Photo credits: (A) Associated Press/Wide World Photos, (B) Joachim Lent.}\label{fig1}
\end{figure}

\begin{figure}[h!]
\begin{center}
\centerline{\includegraphics[width=1.0\textwidth]{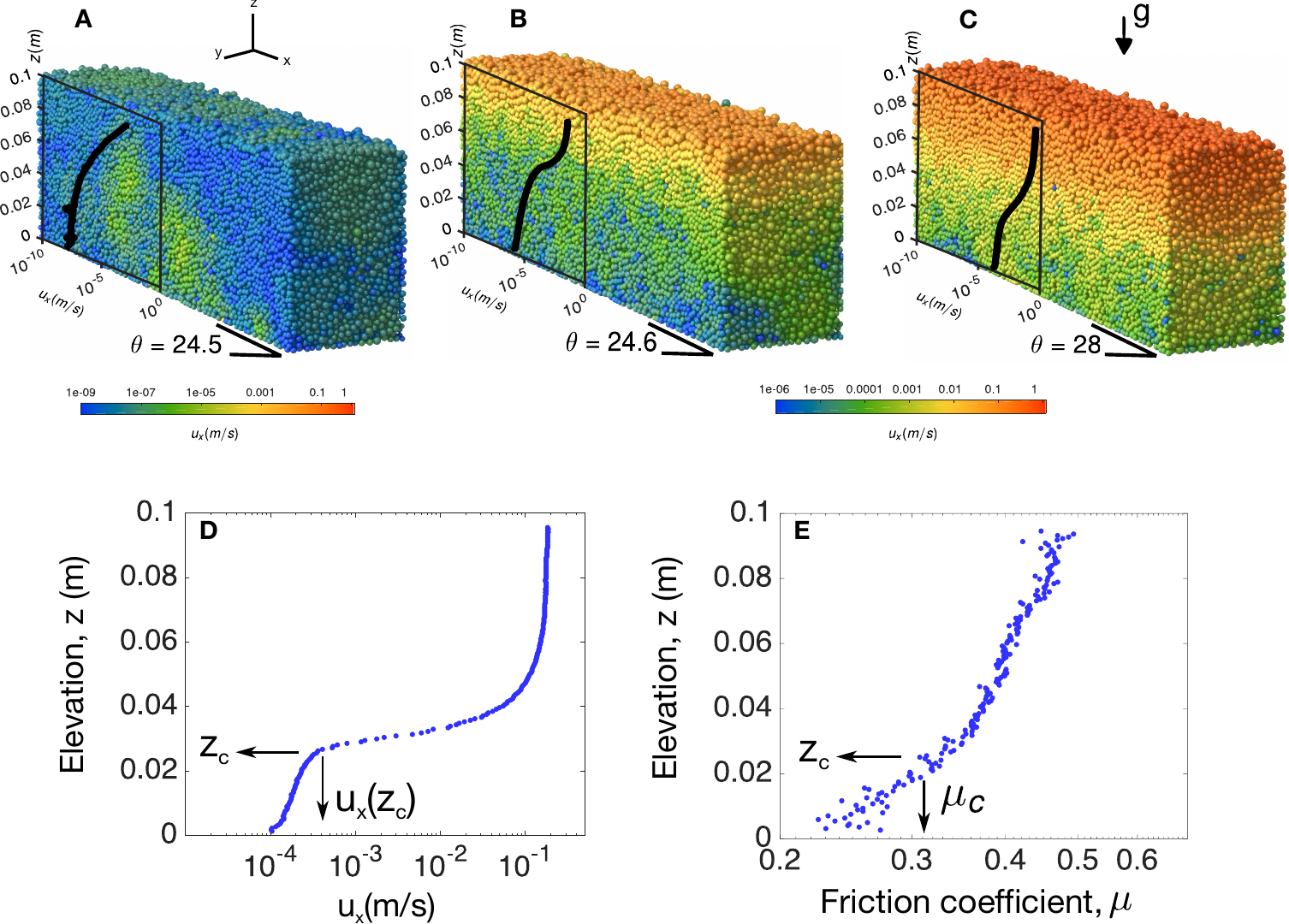}}
\caption{Hillslope DEM simulations. Snapshots (A-C) correspond to time $t=140$ (s) after inclining the granular hillslope; particle colors represent their downslope ($x$-dir) instantaneous velocities, while black lines show time-averaged downslope velocities. The bulk (macroscopic) angle of repose for this set of simulations is $\theta \approx 24.6 \degree$. (A) $\theta = 24.5 \degree$ corresponds to a hillslope just below onset of dense flow at the surface, where the pack is almost fully creeping. (B) $\theta = 24.6 \degree$ is right at the transition point. (C) $\theta = 28 \degree$ shows a fully developed dense granular flow in more than half of the model depth. Panels (D) and (E) show time-averaged downslope ($x$-dir) velocity ($u_x(z)$) and local friction coefficient ($\mu = \tau_{xz} / \sigma_n$) profiles, respectively, for the granular hillslope at $\theta = 28 \degree$. The critical depth $z_c$ and critical downslope velocity $u_x(z_c)$ at the transition to creep are indicated. The value $z_c$ is further used in panel (E) to determine the critical friction coefficient $\mu_c$.}
\label{fig2}
\end{center}
\end{figure}

\noindent Building on the work described above, and recent success in linking idealized granular-physics models to sediment transport \citep{houssais2016toward}, we examine the granular origins of a creep transport equation and the transition to landsliding by developing a granular hillslope model using the Discrete Element Method (DEM) (Fig. \ref{fig2}). Simulations are performed with LAMMPS (\url{http://lammps.sandia.gov}). In typical soil-mantled hillslopes there is a meters-deep, mobile surface layer composed of mostly unbonded particles that are sand-sized and smaller; it is underlain by weathered bedrock (saprolite) composed of increasingly more bonded particles and larger rock fragments, and eventually by unaltered bedrock \citep{anderson2010geomorphology} (Fig. \ref{fig1}). The surface layer may flow as a sheet or as a channel, depending on confinement and material factors. We model an idealized representation of this mobile surface soil; a layer of polydisperse spheres with diameter range $d = [0.0026:0.0042]\,\text{m}$, average diameter $d_{mean} = 0.0033\,\text{m}$ and depth $h = 45 d_{mean}$, over-riding an immobile and frictional bottom boundary (Fig. \ref{fig2}); more details of the model implementation are available in Supplementary Information (SI). The system is inclined at various gradients $\theta_r = 24 \; \text{to} \;29\degree$ that are below, near and above the bulk angle of repose ($\theta_r = 24.6 \degree$; see Methods). From instantaneous particle motions, we compute vertical profiles of time- and horizontal- ($x$) averaged downslope grain velocity, $u_x(z)$, over the duration of each model run (see SI for averaging procedure). 

\begin{figure*}[]
\begin{center}
\centerline{\includegraphics[width=1.0\textwidth]{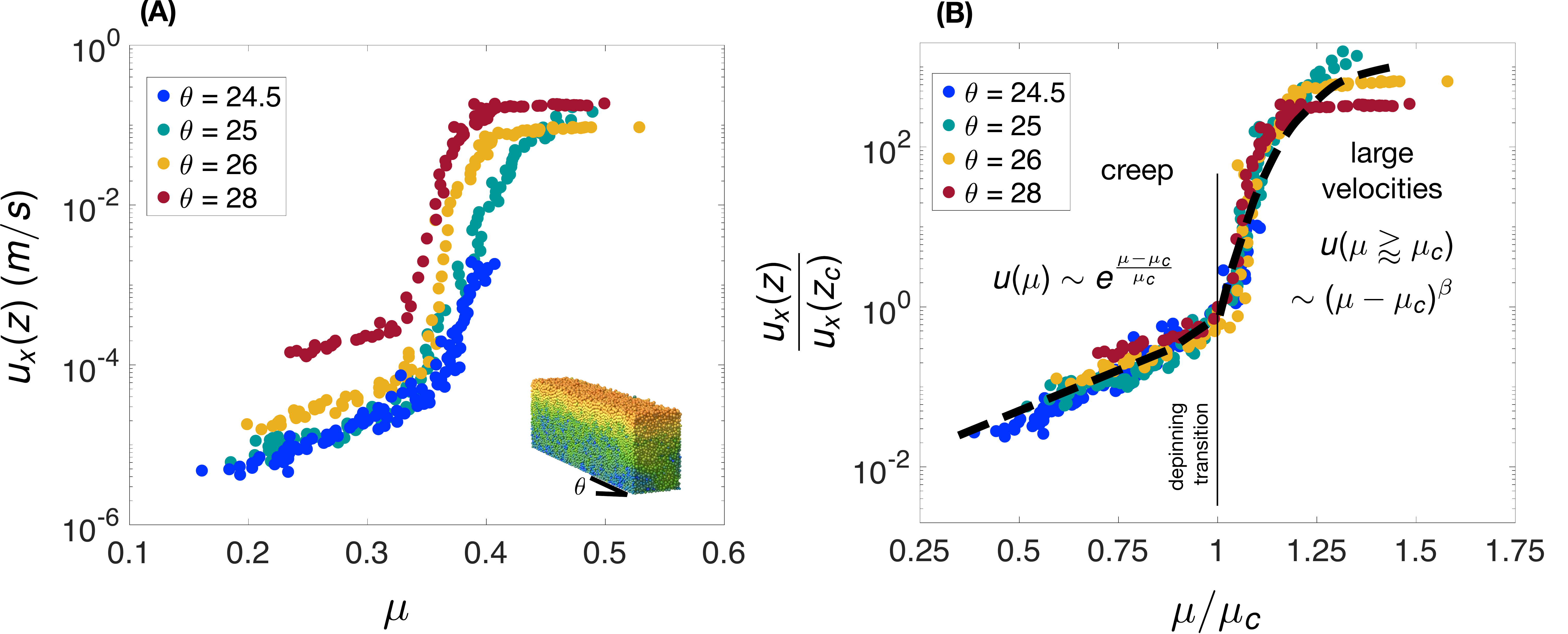}}
\caption{General flow behavior in simulations. (A) DEM results showing local downslope velocity ($u_x(z)$) as a function of local friction coefficient ($\mu$) for four different inclinations, below and above the bulk angle of repose. (B) DEM results showing normalized local downslope velocity ($\frac{u_x(z)}{u_x(z_c)}$) as a function of normalized local friction coefficient ($\frac{\mu}{\mu_c}$) for the four different inclinations shown in panel (A). Dashed line illustrates an exponential scaling for creep regime with critical gradient $\mu_c = 0.3$, and a power-law scaling for the range of large flux with a power-law exponent $\beta = 5/2$; relations are not fit to the data, rather they are shown for illustrative purposes.}\label{fig3}
\end{center}
\end{figure*}

In order to study the dynamical behavior of our model hillslope in the phase transition framework described above, we calculate the per-grain friction coefficient $\mu = \sigma_{xz} / \sigma_{n}$, where $\sigma_{xz}$ is the per-grain stress tensor in the $x-z$ plane and $\sigma_{n} =  \frac{1}{3} (\sigma_{xx} + \sigma_{yy} + \sigma_{zz})$ is the average confining/normal stress (see SI). Similar to velocity, we perform time- and horizontal-averaging to produce vertical profiles of the local friction coefficient for each run; since only the values averaged over time and x-direction are reported, we retain the notation $\mu$ to emphasize that measured friction is local in that it changes with depth. Our simulations show two distinct phenomenological behaviors for inclinations below and above $\theta_r = 24.6 \degree$. Below $\theta_r = 24.6 \degree$ we observe slow particle velocities, a slow decay rate of the downslope particle velocity ($u_x(z)$) with depth, and hot spots of intermittent and localized motion \citep{amon2012hot}. We interpret this regime as creep (Fig. \ref{fig2}A), where similar dynamics have been reported in experiments \citep{houssais2015onset,komatsu2001creep, amon2013experimental}. As $\theta$ crosses $\theta_r$ we observe the abrupt emergence of a fast and continuous surface layer flowing over the creeping regime, whose velocity and thickness increase with increasing $\theta$ above critical (Fig. \ref{fig2}). The rapid decay of $u_x(z)$ with depth, and the continuous nature of the particle motions, are consistent with the dense granular flow regime \citep{houssais2015onset,komatsu2001creep}. We verified that the DEM simulations reproduce the rheology observed in heap-flow experiments that are similar to our model setup (see SI, Fig. S3). The transition from dense-granular flow to creep is associated with a kink in the mean velocity profile, which is used to define a critical depth ($z_c$) and critical particle velocity ($u_x(z_c)$) for each inclination modeled (Fig. \ref{fig2}D). We also observe a kink in the profile of friction coefficient $\mu$ with depth, which occurs at the same critical depth; we infer the associated value as the critical friction associated with the creep transition, $\mu_c$ (Fig. \ref{fig2}E), and compute this also for each inclination. We emphasize here that creeping and dense flow regimes can take place in the same column of soil, with dense flow at the top and creep at the bottom. The variation of downslope particle velocity ($u_x(z)$) versus friction coefficient $\mu$ is qualitatively similar for four different inclinations, below and above the bulk angle of repose (Fig. \ref{fig3}A). These observations, and previous work \citep{yan2016model}, suggest the possibility of a generalized relation between particle velocity and local friction. Comparison of normalized average grain velocity, $u_x/u_x(z_c)$, versus normalized local friction coefficient, $\mu/\mu(z_c)$, for all inclinations confirms this notion and reveals a striking pattern (Fig. \ref{fig3}B). For the creep regime ($\mu/\mu(z_c) < 1$) simulations show an exponential flow relation, $\frac{u_x}{u_x(z_c)}  \propto e^{\frac{\mu-\mu_c}{\mu_c}}$, with a transition friction coefficient of $\mu_c \approx 0.33$ that is similar for all model runs. For the dense granular flow regime ($\mu/\mu(z_c) > 1$) the functional form is a power law, $\frac{u_x}{u_x(z_c)} \propto {(\mu-\mu_c)}^{\beta}$, where $\beta$ is the critical exponent, $\mu(z_c)$ the critical point, and $u_x$ is the order parameter. This is in agreement with the suggestions by Fisher (1985) \citep{fisher1985sliding} and Chauve et al. (2000) \citep{chauve2000creep} that the pinned to sliding transition is a second-order phase transition in which the order parameter obeys power-law scaling close to the critical point. A similar depinning transition with an exponential law relation at a low driving force and a power law relation above a critical threshold has been also reported in failure of inhomogeneous brittle materials \citep{ponson2009depinning}. For a plastic depinning transition it is expected that the critical exponent $\beta > 1$ \citep{reichhardt2016depinning}, consistent with our simulation results. A precise value for the exponent cannot be determined from numerical results alone. Moreover, finite-size effects and dimensionality of our system may influence the value of the critical exponent. Nevertheless, a representative value $\beta = 5/2$, shown for illustration purposes (Fig. \ref{fig3}B), is in the general range reported for colloidal and granular systems near the critical point \cite{habdas2004forced,lajeunesse2010bed,houssais2015onset}. 

\section*{\hspace{-0.2in}Hillslope evolution as a depinning transition: field evidence}
\noindent We acknowledge that our highly idealized simulations may not translate to downslope movement of heterogeneous soil in the complex natural environment. The above findings indicate, however, that hillslope soil movement may be governed --- at least in part --- by generic glassy dynamics associated with disordered granular systems. To search for signatures of this behavior in the field, we have collected observations of sediment flux ($q_s$) as a function of local hillslope gradient from five different published sources with different climatic and uplift conditions, as well as different soil types and material properties (Table S1). Unfortunately, data collection methods and their associated timescales differ among these studies, potentially contributing to noise and limiting our ability to directly compare among these different field sites. The data presented by Yoo et al. (2007) \citep{yoo2007integration} are from Frog's Hollow, a semiarid eucalyptus grassland savannah hillslope located about 80 km south?southeast from Canberra, New South Wales, Australia. They calculated sediment transport rates by taking soil samples along a hillslope transect, measuring the soil mass production rate and the elemental chemistry of soils and saprolite, and then using an iterative modeling process integrating over the timescales of chemical weathering. Data presented by Gabet (2003) \citep{gabet2003sediment} are from the Santa Ynez Valley in the tectonically active transverse ranges near Santa Barbara, U.S. state of California, with a semiarid Mediterranean climate. The measurements were carried out over annual timescales using sediment traps installed on hillslopes. Data by Martin \& Church (1997) \citep{martin1997diffusion} and Martin (2000) \citep{martin2000modelling} are from the Queen Charlotte Islands, off the coast of British Columbia in Canada, with an oceanic climate and extremely frequent precipitation. The original datasets in these studies are from reports by Rood (1984) \citep{rood1984aerial} and Rood (1989) \citep{rood1989site}, which are landslide inventories completed by identifying landslides on aerial photographs in the region. They cover approximately a 40-year period of activity. 

\begin{figure*}[]
\begin{center}
\centerline{\includegraphics[width=1.0\textwidth]{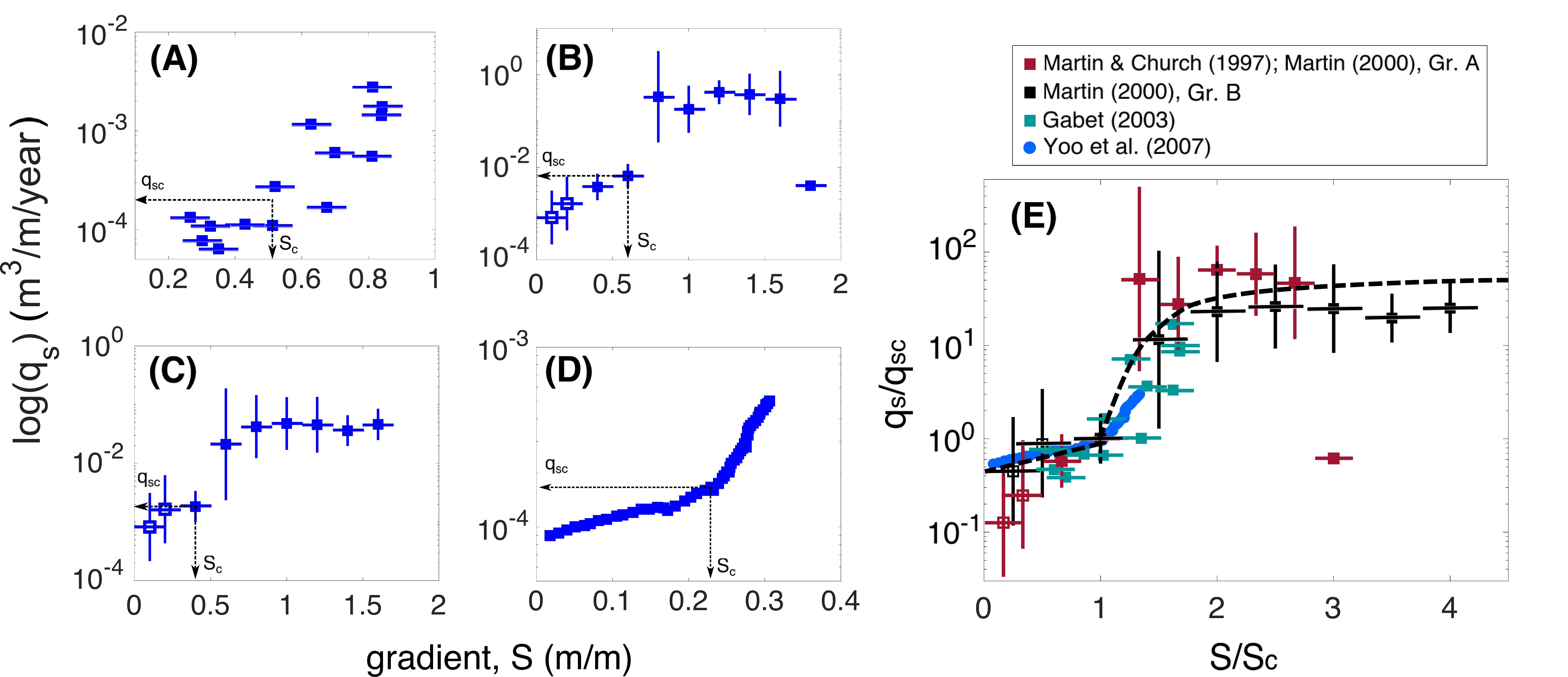}}
\caption{Field data showing measured sediment flux $q_s$ versus hillslope gradient $S$ for natural hillslopes reported previously in the literature. (A) Gabet et al. (2003) \citep{gabet2003sediment}, (B) Group A basins in Martin (2000) \citep{martin2000modelling} and Martin \& Church (1997) \citep{martin1997diffusion}, (C) Group B basins in Martin (2000) \citep{martin2000modelling}, and (D) Yoo et al. (2007) \citep{yoo2007integration}. (E) Rescaled data combined from panels (A)-(D). Note that normalized flux $q_s/q_{sc}$ is equivalent to normalized velocity while normalized gradient ($S/S_c$) is equivalent to normalized friction, allowing comparison to numerical results (Fig. \ref{fig3}B). The dashed line in panel (E) shows the same bi-partite flux relation as figure \ref{fig3}B for illustration purposes; i.e., an  exponential flux relation for gradients below critical, and a power-law relation for larger gradients.}\label{fig4}
\end{center}
\end{figure*}

In studies by Yoo et al. (2007) \citep{yoo2007integration} and Gabet et al. (2003) \citep{gabet2003sediment}, the sediment flux was originally measured in units of mass flux each for basin. The averaged bulk density of soil in the study by Yoo et al. (2007) is reported as 1800 $kg/m^3$ \citep{yoo2007integration}, and in the study area by Gabet et al. (2003) is reported as 1770 $kg/m^3$  \citep{gabet2003sediment}. We used these densities to convert mass sediment flux to volumetric sediment flux (Figs. \ref{fig4} A \& D). The studies by Martin (2000), Martin \& Church (1997) originally reported measurements of volumetric sediment flux. They presented measurements for twenty three different basins: thirteen sites (Group A) have the majority of their basins composed of soft volcanic and sedimentary rocks, while the ten remaining sites (Group B) have a larger proportion of their basins composed of hard volcanic rocks and granites. Here, we analyze the data from their Group A and Group B hillslopes separately in order to calculate critical slope and flux at that critical slope. We further calculated the mean value of sediment flux measurements for each hillslope gradient class in each group; these are presented in a single sediment flux versus hillslope gradient relationship for the studied region (Figs. \ref{fig4} B \& C). We assumed a constant errorbar of 0.1 for all gradient measurements (classes) in the studies by Martin (2000) and Martin \& Church (1997), even though such an error in measurements is not explicitly stated in their studies. All data here were extracted from figures in the cited papers and also reports cited therein. For the case of studies by Martin (2000) and Martin \& Church (1997), the values of sediment flux and their errorbars at slopes smaller than 20\degree \, are calculated based on the estimates presented in Table 1 in Martin \& Church (1997). 

For each study we observe a kink in the relation between flux and slope, which allows us to determine critical values of $q_{sc}$ and $S_c$ for each field site by eye from inflection points in the plots (Table S1). These critical values are illustrated with arrows in Figs. \ref{fig4} A-D. The normalized sediment flux and the normalized hillslope gradient are then calculated as $q_s/q_{sc}$ and $S/S_c$, respectively. Note that sediment flux $q_s = \langle u_x(z) \rangle h$, where the brackets indicate averaging over the depth of the soil column ($h$). While studies report values for $q_s$, most are derived from surface measurements and therefore mostly reflect surface velocities of the flows, while flow depths are generally poorly constrained. Normalization removes this depth dependence, however, since $q_s/q_{sc} = u_x(z)/u_x(z_c)$ Hillslope gradient is also related to friction coefficient; assuming a naive hydrostatic and 1D behavior for a depth-averaged earth flow, $\sigma_{xz} = \mu_{soil} \sigma_n \rightarrow \rho g h S = \mu_{soil} \rho g h \rightarrow S \sim \mu_{soil}$, where $g$ is acceleration due to gravity. Thus, data plots of normalized flux and slope for the field data are equivalent to normalized velocity and friction presented from numerical simulations.

The critical values for slope and flux likely encode climatic, tectonic and soil properties unique to each site \cite{perron2017climate}. Plotting normalized values from all field sites as $q_s / q_{sc}$ against $S/S_c$, however, collapses the data and results in a pattern that is similar to our model results of $u_x/u_x(z_c)$ against $\mu/\mu(z_c)$ (Fig. \ref{fig4}E). We take this similarity as strong evidence for a generic depinning transition, where field data are consistent with an exponential flux relation for gradients below critical and a power-law relation for larger gradients. Moreover, the values inferred for critical slopes at each field site are physically meaningful; they correspond to a reasonable range of reported values for the angle of repose (or friction coefficient) of soils (Table S1) \citep{fall2014sliding,skempton1985residual}. The scatter in the data, especially above the critical gradient, can be due to several factors including: (i) limited sediment availability on natural hillslopes, because high slopes transport soil faster than it can be produced from weathered bedrock; (ii) different measurement methods used in different studies, including variations in the detection limits for transport.

Some additional features of the data warrant mention, in terms of their physical interpretation. Hillslope creep velocities (fluxes) are measured over years to decades, and thus average over event-based and seasonal fluctuations in flow speed that often occur due to precipitation and temperature effects \citep{saunders1983rates,martin1997diffusion}, bioturbation \citep{gabet2000gopher} and other disturbances. The timescales associated with measured velocities (fluxes) for above-critical flows are also longer than that of the individual landslides that (presumably) occur. Thus, the reported velocities (fluxes) of soil motion should be understood as the long-term average of episodic and relatively fast events, and intervening periods of relatively slow motion. This is not unlike mountain uplift that results from repeated fault slip; average uplift rates are not representative of slip events, but are nonetheless meaningful for considering landscape erosion that occurs over geologic timescales \citep{dietrich2003geomorphic}. Another notable feature is that fluxes vary widely for slopes slightly to moderately larger than critical ($1 < S/S_c < 1.5$). We speculate that flows within this slope range may occur as either creep or landslides, depending on environmental forcing (e.g., pore pressure \citep{iverson1997physics}), supply (or availability) of material, and soil thickness. As a result, the creep-to-landsliding transition in natural landscapes is much more variable than in a constant forcing situation such as our model. We suspect that averaging over suitably large (geologic) timescales, if possible, would recover a flux-slope relation that is similar to model expectations but with a more diffuse flow transition. Such an idea may be tested by examining the topographic hillslope profile produced by erosion and rock uplift over geologic timescales \citep{roering1999evidence}.\\

\section*{\hspace{-0.2in}Modeling landscape evolution with a ``glassy flux model''}
\noindent We propose a bi-partite ``glassy flux model'' to represent hillslope soil transport, that joins the exponential and power-law relations associated with creep and landsliding regimes, respectively: 

\begin{equation} 
q_s/q_{sc} =  e^{\frac{S-S_c}{S_c}}\mathcal H(S_c-S)+[A\,(S-S_c)^\beta + 1] \mathcal H(S-S_c)
\end{equation}

where $\mathcal H$ is the Heaviside step function that acts to blend the two transport regimes across the transition  \citep{chauve2000creep}, and $A$ is a constant associated with a particular field site. This equation implicitly assumes steady flow, and therefore is only applicable for long timescales that integrate over very many flow events (see SI). The flux equation (1) is related to hillslope erosion through conservation of mass:

\begin{equation} 
-\rho_s \frac{\partial z}{\partial t} = \rho_s \nabla \cdot q_s + \rho_r C_o,
\end{equation}
 
where $\rho_s$ and $\rho_r$ are the bulk densities of sediment and rock, respectively, $\partial z/ \partial t$ is the rate of landscape elevation change, and $C_o$ is the rock uplift rate. We use equations (1) and (2) to model the steady-state form of a hillslope, i.e., the topography associated with a balance between uplift and erosion (see Methods) that results from the new flux equation. The left boundary condition of the model hillslope is no flux, representing a drainage divide, and the right boundary condition is a fixed elevation that represents baselevel. We calibrate and compare model results to hillslope topography data in the Oregon Coast Range (OCR). First we assume a value for the critical exponent $\beta = 5/2$ for simplicity because it cannot be better constrained from the data, and then estimate the values for critical gradient $S_c \approx 0.5$
and $A =222$ from values of sediment flux versus gradient reported by Roering et al. \citep{roering1999evidence} from a site near Coos Bay (see SI, Figure S5). Next, we model the transient evolution of hillslope topography by iteratively solving equations (1) and (2) starting from a flat initial condition with constant uplift rate $C_o = 0.075 \, \text{mm/yr}$ and densities $\rho_r/\rho_s = 2$ determined from Roering et al. \citep{roering1999evidence} for OCR. The profile is evolved for 20 million years, roughly the time that OCR has been uplifting \citep{orr1992geology}, and we verify that the hillslope reaches a steady state topography over this time.

\begin{figure}[ht]
\begin{center}
\centerline{\includegraphics[width=1.0\textwidth]{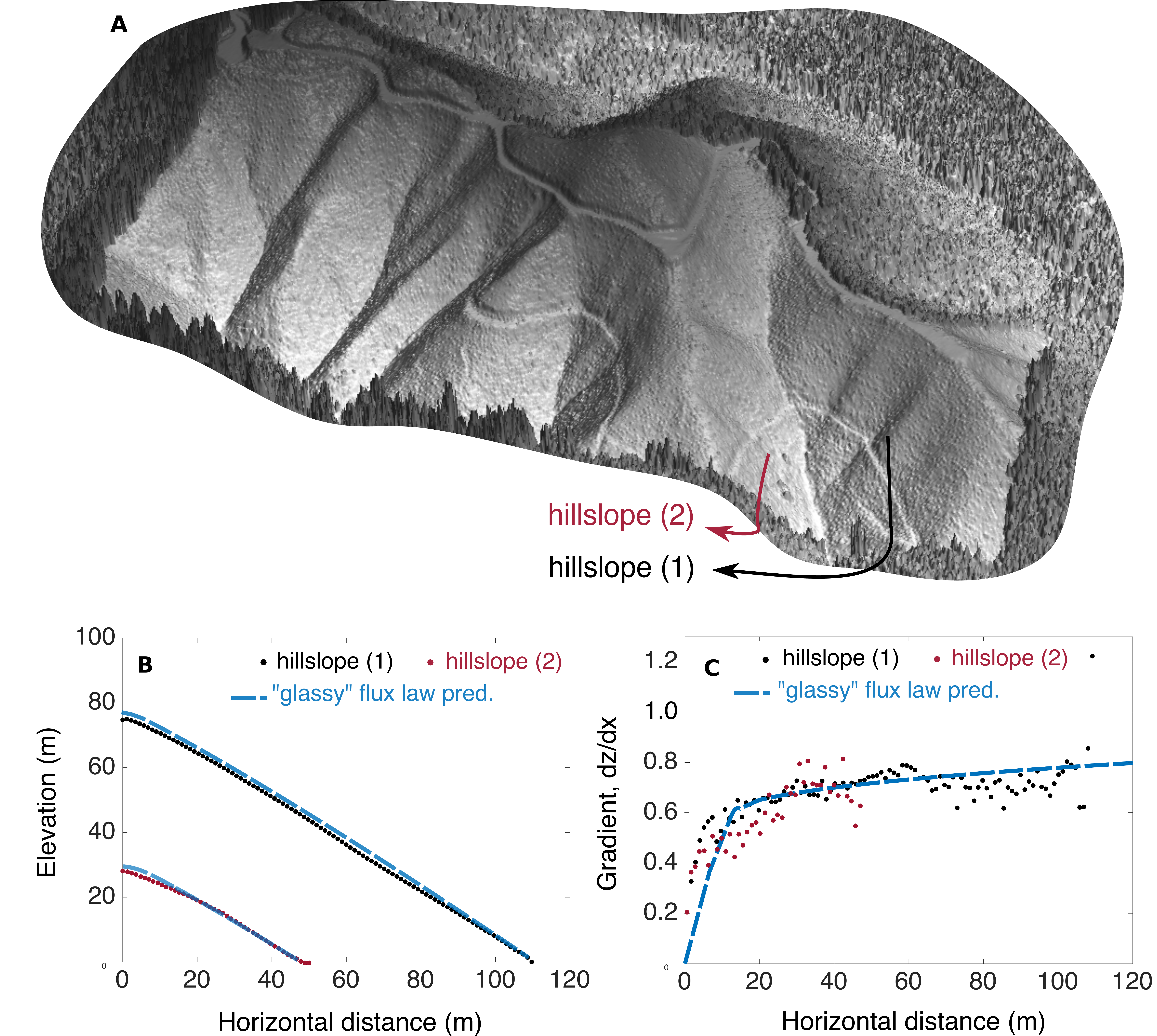}}
\caption{Hillslope topography of the Oregon Coast Range (OCR) derived from publicly available airborne lidar data \cite{notemapfig5}. (A) Regional perspective view, showing locations of two example hillslopes. (B) The elevation-distance and (C) gradient-distance relationships for representative profiles of hillslopes (1) (black dots) and (2) (red dots) in panel (A). Blue dashed line is the prediction of the ``glassy'' flux model with $S_c = 0.5$ and $\beta = 5/2$.  See SI for more examples.}
\label{fig5}
\end{center}
\end{figure}

We compare model results to hillslope topographic profiles extracted from aerial LiDAR data at an OCR site 60 km West of Eugene, Oregon (see SI). The glassy flux model reasonably captures elevation and gradient profiles for both short hillslopes (distance about 40 m) where gradients are mostly below critical, and longer hillslopes which significantly exceed the critical slope (Fig. \ref{fig5}). Hillslope gradient profiles show a clear kink at the critical slope value $S_c \approx 0.5$ derived from the glassy flux model (Fig. \ref{fig5}C); the corresponding angle of 26 degrees represents a reasonable value for the transition from creep to landsliding. The explicit incorporation of landslide (dense-granular flow) dynamics allows the glassy flux model to reproduce the flattening out of hillslope profiles as they lengthen; this flattening has been previously reported, and cannot be reproduced with diffusion-like flux equations \citep{grieve2016long} (Fig. S8). We also verified that the model reproduces observed hillslope topography in a different climatic and geologic setting in California (see Fig. S7).

\section*{\hspace{-0.2in}Discussion}
\noindent We have developed a model for hillslope soil transport that, for the first time, describes behaviors from creep and slow-earthflow to landsliding and fast-flow regimes. Although the DEM simulations are highly idealized, we suggest that the underlying dynamics are general. While soil creep on hillsides has been viewed as the result of external perturbations \citep{culling1963soil,roering1999evidence,gabet2000gopher}, model results show that this is not necessary. We found that the addition of perturbations, through random noise added to the locations of some grains, increased the flux magnitude in the creeping regime but did not change the functional form of the flux-slope relation (Fig. S2). Noise had little influence on the fast flow regime (Fig. S2). We also confirmed that changing grain shape does not change the qualitative behavior in simulations, by replacing spheres with elongated particles having a 3:1 aspect ratio (Fig. S4). More broadly, observations provide strong evidence that the creep-landslide transition exhibits glassy dynamics, that may be modeled as a plastic depinning transition at a critical normalized force. The sub-critical exponential relation, and the super-critical power-law scaling (equ. 2), are expected behaviors based on theoretical and experimental studies of amorphous systems \citep{reichhardt2016depinning}. In other words, such behavior is the generic consequence of dynamical phase transitions in disordered materials. A dimensionless number, local normalized friction coefficient in the numerical model, $\mu/\mu_c$, is calculated in terms of the effective tangential and normal stresses. Although the current model does not account for changes in fluid pore pressure that are known to influence downslope flow velocity \citep{iverson1997physics}, we suggest this effect might be viewed as a perturbation to the effective pressure terms that could be combined in the future with the framework we propose. The effective critical friction coefficients determined for the different field settings examined here fall in the range expected for soil mixtures \citep{fall2014sliding,rowe1969relation}. 

These results show how recent advances in the physics of disordered materials can be used to explain the evolution of natural landscapes over geologic timecales. The functional form of the flux equation $q = f(\mu)$ used in this work is a specific case of a more general form $q = f(\mu,\eta)$, where $\eta$ represents mechanical internal and external noise \cite{DeGiuli15082017}. We suggest that creep, and its associated slow subcritical flow, takes place in our numerical system and in natural hillslopes due to: (i) internal disorder of the particulate packing; and (ii) bedrock and saprolite boundary layers that surround the mobile regolith, which continuously inject disorder that may induce creep through non-local effects \citep{pouliquen2009non,bouzid2013nonlocal,kamrin2015nonlocal}. It is an open question for soil-mantled hillslopes whether, and under what conditions, the injection of porosity and noise from external perturbations (plants/animals, freeze/thaw/swell, etc.) produces distinctly different dynamics from the sources of disorder considered here.

Finally, we note that one of the hallmarks of granular and amorphous materials is the emergence of rate weakening in the vicinity of their dynamical phase transition \cite{richard2008rheology,dijksman2011jamming}. Although the fundamental mechanisms of this phenomenon remain to be explored, we believe the picture provided here can help to understand the origins of rate- and state-dependent friction behavior that has recently been proposed to characterize slow and fast landslides \cite{handwerger2016rate,lucas2014frictional}.\\

\textbf{\hspace{-0.2in}Description of supplementary materials} 
Details of the implementation of the DEM model are described in SI section 1 and Table S2, and the protocol for calculation of velocity and stress profiles from DEM simulations are provided in SI Section 2. The influence of perturbation on the behavior of DEM model in presented in SI Section 3. We compare the observations from our DEM simulations with a local granular rheology model applied to a heap flow granular experiment in SI section 4. The influences of grain shape on slow creep regime in DEM simulation and experiments are discussed in SI section 5. Implementation of the landscape evolution model is described in SI section 6. Details for measurements of hillslope profiles from lidar data are described in SI section 7.\\
\textbf{\hspace{-0.2in}Acknowledgements} Research was supported by US Army Research Office, Division of Earth Materials and Processes grant 64455EV, US National Science Foundation (NSF) grant EAR-1224943, NSF INSPIRE/EAR-1344280, and NSF MRSEC/DMR-1120901. BF was a synthesis postdoctoral fellow of the National Center for Earth-surface Dynamics (NCED2 NSF EAR-1246761) when this work was performed. B.F. also acknowledges support from the Department of Geosciences, Princeton University, in form of a Hess Fellowship. Lidar data acquisition and processing completed by the National Center for Airborne Laser Mapping (NCALM - http://www.ncalm.org). We thank J. Roering, S. Mudd, T. Perron and J. Prancevic for comments and discussions that improved this manuscript, and J. Roering for providing data from the OCR. The authors declare that they have no competing financial interests.\\ 
\textbf{Author Contributions} DJJ, BF and CPO designed the study, and all discussed the research steps and results to shape this manuscript. All authors contributed to writing of the manuscript. BF designed and ran DEM and continuum (landscape evolution) simulations and compiled the field data. BF and CPO analyzed simulation results and field observations. BF and CPO prepared figures.

%

\newpage

\section*{\hspace{-0.2in} Supplementary Information}
\noindent
\vspace{-0.7in}
\setcounter{figure}{0}
\renewcommand{\thefigure}{S\arabic{figure}} 
\renewcommand{\thesubsection}{\arabic{subsection}}   
\renewcommand{\thetable}{S\arabic{table}}  

\section{Implementation of the DEM model}

\noindent The Discrete Element Method (DEM) model consists of a packing of 54780 polydisperse grains with a diameter range $d = [0.0026:0.0042]\,\text{m}$ and average diameter size, $d_{mean} = 0.0033\,\text{m}$. Grain density and contact Young's modulus are chosen equal to properties of glass beads (Table S2). The model domain (Fig. 2) is rectangular with a periodic boundary condition applied in the downslope ($x$) direction. The size of the system in each direction is $L_x = 5L_y = 2L_z=90 \;d_{mean}$. The system is initially prepared by randomly inserting and raining (under gravity) grains in the simulation box with a desired initial packing fraction, $\nu \sim 0.4$. The system is then allowed to relax for 500 million time-steps ($35\,\text{s}$) and to achieve a packing fraction of $\nu = 0.5$, before inclining it at various gradients below, near and above the bulk angle of repose of the granular system ($\theta_r = 24.6 \degree$;). The restitution coefficient is chosen to be very small ($e_n = 0.01$ for Stokes number $< 1$ , ref. \cite{gondret2002bouncing}) such that collisions are highly damped (Table S2), as is the case in natural soil-mantled hillslopes. The grains are modeled as compressible spheres of diameter $d_{s,l}$ that interact when in contact via the Hertz-Mindlin model \cite{johnson1987contact,landau1959theory,mindlin1949compliance}:

\begin{equation} \label{eq:sg16}
F=(k_n \delta \vec{n}_{ij}- \gamma_n \vec{v}\vec{n}_{ij})+(k_t \delta \vec{t}_{ij} - \gamma_t \vec{v}\vec{t}_{ij})
\end{equation}

where the first term is total normal force, $\vec{F}_n$, and the second term is total tangential force, $\vec{F}_t$. In Equation~\ref{eq:sg16}, $k_n$ and $k_t$ are normal and tangential stiffness respectively, $\delta$ is the overlap between grains, $\gamma_n$ is the normal damping, $v$ is the relative grain velocity, $\vec{n}_{ij}$ is the normal vector at grain contact, $\vec{t}_{ij}$ is the tangential vector at grain contact, and $\gamma_t$ is the tangential damping. The full model implementation is available on the LAMMPS webpage and several references \cite{zhang2005jamming,silbert2001granular,brilliantov1996model}. The DEM model system is frictional, meaning that the coefficient of friction, $\mu$, is the upper limit of the tangential force through the Coulomb criterion $F_t = \mu F_n$. The tangential force between two grains grows according to non-linear Hertz-Mindlin contact law until  $ F_t/ F_n = \mu$ and is then held at $F_t = \mu F_n$  until the grains lose contact. The damping coefficients $\gamma_n$ and $\gamma_t$ are determined within the implementation of LAMMPS from the chosen value for the restitution coefficient, $e_n$. 

\section{Calculation of velocity and stress profiles in DEM}

\noindent The grain horizontal velocity is measured by tracking individual particles for the full duration of each simulation, and also accounting for the periodic boundary condition in the horizontal (x-) direction. The simulations reported in the main manuscript are run for a duration of $\sim$1 minute ($\sim$900 million time-steps). The grain velocities are time-averaged during the final 1 million timesteps of each run. This time duration is enough for all grains to displace at least one average grain diameter in all runs. The time-averaged grain velocity is then horizontally averaged by dividing the full depth of the granular layer into 200 bins. The velocity profiles presented here and in the main paper, including those used for studying the phase transition, are all time- and horizontally (x-dir) averaged in the same manner. 

We checked the influences of the duration of simulations on the dynamics of the system by running simulations --- at three inclinations, below and above the bulk angle of repose --- for ten times longer (total duration $\sim$10 minutes $\sim$9 billion time-steps) than those reported in the main paper. The velocity profiles for those simulations, also time-averaged during the final 1 million timesteps at each run, are shown in Fig. S1 and are indicative of no qualitative change in the dynamics of the system as time goes on. 

The per-grain stress tensor was computed using the general Virial function \cite{thompson2009general}. We computed friction coefficient profiles for each run using the same time- and horizontal-averaging procedure as was done for velocity profiles. 

\section{Influence of perturbations}

\noindent We run perturbed simulations to explore the influence of external disturbances on slow and fast flow dynamics. In simulations with external perturbations, they are applied for the full duration of the run to grains as a randomly distributed displacement within the amplitude range [-A,A], in all directions. A maximum disturbance amplitude of $A = 0.005 \times$average grain diameter [$A = 1\times 10^{-5} m$] is used. In Figure S2, we compare the variation of normalized local horizontal velocity ($\frac{u_x(z)}{u_x(z_c)}$) as a function of normalized local friction coefficient ($\frac{\mu}{\mu_c}$) in non-perturbed and perturbed runs at inclinations below and above the bulk angle of repose. Simulations with perturbations $A = 1\times 10^{-5} m$ show an elevated magnitude of rearrangement and horizontal velocity in the creep/slow dynamics regime. All other aspects, however, are not affected qualitatively by perturbations: the regime of fast dynamics and large velocities, the form of the transition from slow to fast regimes, and the critical value of the friction coefficient at which the transition takes place. A more rigorous study is required to investigate the influences of perturbation amplitude and frequency, and their general control over slow and fast regimes.

\section{Granular rheology}

\noindent We compared the rheology of the granular layer in our simulations with the indirect measurements of granular rheology in heap-flow experiments previously performed by Richard et al. (2008) \cite{richard2008rheology}. This is done by investigating the relationship between local friction coefficient and local inertial number in the numerical model and experiments. The inertial number is defined as \cite{jop2006constitutive}

\begin{equation} 
I_n = \frac{\dot\gamma \, d_{mean}}{\sqrt{\sigma_n/ \rho}}
\end{equation} 

where $\dot\gamma$ is the granular shear rate, $d_{mean}$ is the average grain diameter, $\sigma_n$ is the confining (normal) pressure, and $ \rho$ is the grain density. The information needed to calculate local inertial number is readily available from previously described calculations of the velocity and stress profiles in the simulations. In the experiments, the measurements of local shear and normal stresses and local longitudinal velocity have been performed by Richard et al. \cite{richard2008rheology}. They used momentum equations for steady, fully developed heap-flows of inclined glass bead layers, combined with local 3D volume fraction calculations from numerical simulations to derive stress balance equations and ultimately normal and shear stress profiles. Their measurements have been performed at different degrees of inclination (35, 45, and 55 degrees). We used their measurements to calculate local friction coefficient and local inertial number for all three tested inclinations in their study. The friction coefficient measurements are normalized by the value of friction coefficient at the transition to creep in their measurements. Likewise, friction coefficient calculations in our DEM simulations are normalized by the value of friction coefficient at the transition point. The calculations of local friction coefficient and local inertial number in our DEM simulations are performed for inclinations, 24.6, 26, 27 and 29 degrees. The result is presented in Figure S3, and suggests a good agreement between rheology of the granular layer in our numerical system and the experimental heap-flow by Richard et al. \cite{richard2008rheology}.

\section{Influence of grain shape on slow creep}

\noindent We have run simulations with elongated grains to explore whether creep persists in our DEM model with non-spherical grain shapes. Particles are made of three rigidly attached spherical grains (each grain diameter, $d = 0.001$ m, $d_{long,elongated} = 0.003$ m) with the same mechanical properties as the spherical grains reported above. The simulations are run by continuously pouring grains at a rate of 2.5 $\frac{grain}{s}$ from a height $H = 40 \times d_{long,elongated}$ and a container of size $[l_x,l_y,l_z] = [10,5,1]\times d_{long,elongated}$ on a flat frictional surface/wall and allowing them to form a symmetric hill with an almost steady-state shape. There are frictional sidewalls in the y-dir, whereas the x-dir boundaries of the system are open, meaning particles that reach the x-dir limits leave the system. The system size is $L_x = 12.5L_y = 2L_z=85 \;d_{long,elongated}$.  Figure S4-A shows a visualization of the numerical setup with grains colored according to the magnitude of their instantaneous horizontal velocity. The visualization shows the flow and rearrangements not only on the surface of the granular heap in the form of a dense rapid flow, but also in the subsurface in the form of slow creep.

Komatsu et al. (2001) have previously studied creep and dense granular regimes in a heap flow experimental setup similar to that shown in Fig. S3-B, using spherical, elongated and angular particles \cite{komatsu2001creep}. The velocity profiles in their experiments are shown in Fig. S4-B and suggest the persistence of creep motion with different grain shapes (Figs. S4-D to 4-F). Their observations indicate a higher decay rate in the exponential velocity profile of angular and elongated grains, compared to spherical particles, but qualitatively similar behavior.  

\section{Implementation of the landscape evolution model}

\noindent We model the transient evolution of hillslope topography by iteratively solving the mass continuity equation for the duration of uplift. Mass continuity/conservation can be written as:

\begin{equation} 
-\rho_s \frac{\partial z}{\partial t} = \rho_s \nabla \cdot q_s + \rho_r C_o,
\end{equation}
 
where $\rho_s$ and $\rho_r$ are the bulk densities of sediment and rock, respectively, $\partial z/ \partial t$ is the rate of landscape elevation change, $C_o$ is the uplift rate, and $q_s$ is the volumetric sediment flux. Equation (3) is solved using a finite volume numerical scheme developed originally in the CHILD landscape evolution model\cite{tucker19993d}. The full domain size (length of hillslope) is divided into evenly spaced discrete parcels/cells. We use 40 cells for all simulations. The time step size is defined based on a form of linearization using the Courant condition \cite{courant1967partial}. The left boundary condition of the model hillslope is no flux, representing a drainage divide, and the right boundary condition is a fixed elevation that represents base level. 

\section{Measuring hillslope profiles from lidar data}

\noindent For the analysis of hillslope topography in the Oregon Coast Range (OCR) presented in Figure 5 and Figure S8, we used the airborne lidar data collected through an NCALM Seed grant (PI: Kristin Sweeney) that is available via opentopoID: OTLAS.052014.26910.3 on the opentopography platform. We used the Triangulated Irregular Network (TIN) generated DEM images (GeoTiff) directly processed on the opentopography platform. The default grid resolution of 1 meter and maximum triangle size of 50 units have been used for processing purposes. The hilleslope profiles are then measured from DEM images using the Quantum GIS open-source package \cite{QGIS_software}. The measurements are averaged over ten equally-spaced profiles for each hillslope. 

\section{Collections of sediment flux and hillslope surface gradient in the study by Roering et al. (1999)}

\noindent The data presented by Roering et al. (1999) \cite{roering1999evidence} belong to the Oregon Coastal Range in the U.S. state of Oregon, with a mild maritime climate. These data are not actually measurements of flux but rather are modeled from inversions of topography assuming that erosion balances uplift. This study originally reported estimates of volumetric sediment flux and included data for five basins in one region. However, we could only obtain data for one of the basins from one of the authors (J. Roering) (Fig. S5), and the other basins data were not accessible. For this study, we found the value of critical gradient ($S_c$) by finding the range of gradient ($S$) where an exponential relationship best fit the variation of sediment-flux versus hillslope gradient measurements ($S_c \sim 0.487$). The corresponding value of sediment flux represents flux at the transition point ($q_{sc}$). These are shown with arrows in Fig. S5.

\newpage

\clearpage
\noindent {Table S1.} Critical values of gradient and flux ($S_c$ and $q_{sc}$, respectively) for four sources that are used for the field data compilations.\\

{\center \small
\begin{tabular}{l*{6}{c}r}
Source              & \specialcell{Yoo et al.\\ (2007) \cite{yoo2007integration}} & \specialcell{Gabet et al.\\ (2003) \cite{gabet2003sediment}} & \specialcell{Group A hillslopes \\ in Martin (2000) \cite{martin2000modelling}\\ Martin \& Church (1997) \cite{martin1997diffusion}} &  \specialcell{Group B hillslopes \\ in Martin (2000) \cite{martin2000modelling}} & \specialcell{Roering et al.\\ (1999) \cite{roering1999evidence}} \\
\hline
$S_c$ & 0.229 & 0.513 & 0.6 & 0.4 & 0.487 \\
$q_{sc} \, \text{(m$^3$/m/yr})$  & $1.66 \times 10^{-4}$ & $1.11 \times 10^{-4}$ & $6.5 \times 10^{-3}$ & $1.23  \times 10^{-3}$ & $2.0 \times 10^{-3}$  \\
\end{tabular}
}\\
\clearpage

\noindent {Table S2.} DEM simulation parameters\\
\begin{tabular}{l r}
Parameter              & Value \\
\hline
Grain density, $\rho$ & 2500 $\text{kg}/\text{m}^\text{3}$  \\
Gravitational acceleration, $\vec{g}$  & 9.81 $\text{m}/\text{s}^\text{2}$  \\
Young's modulus, $E$     & $50\times 10^{9}$ $\text{N}/\text{m}^\text{2}$  \\
Poisson ratio, $\nu$    & 0.45  \\
Friction coefficient, $\mu$    & 0.5  \\
Coefficient of restitution, $e_n$     & 0.01  \\
Time step, $\Delta t$     & $7\times10^{-8}$ s  \\\\
\label{tab1}
\end{tabular}

\clearpage

\begin{figure}
  \centering
      \includegraphics[width=1.0\textwidth]{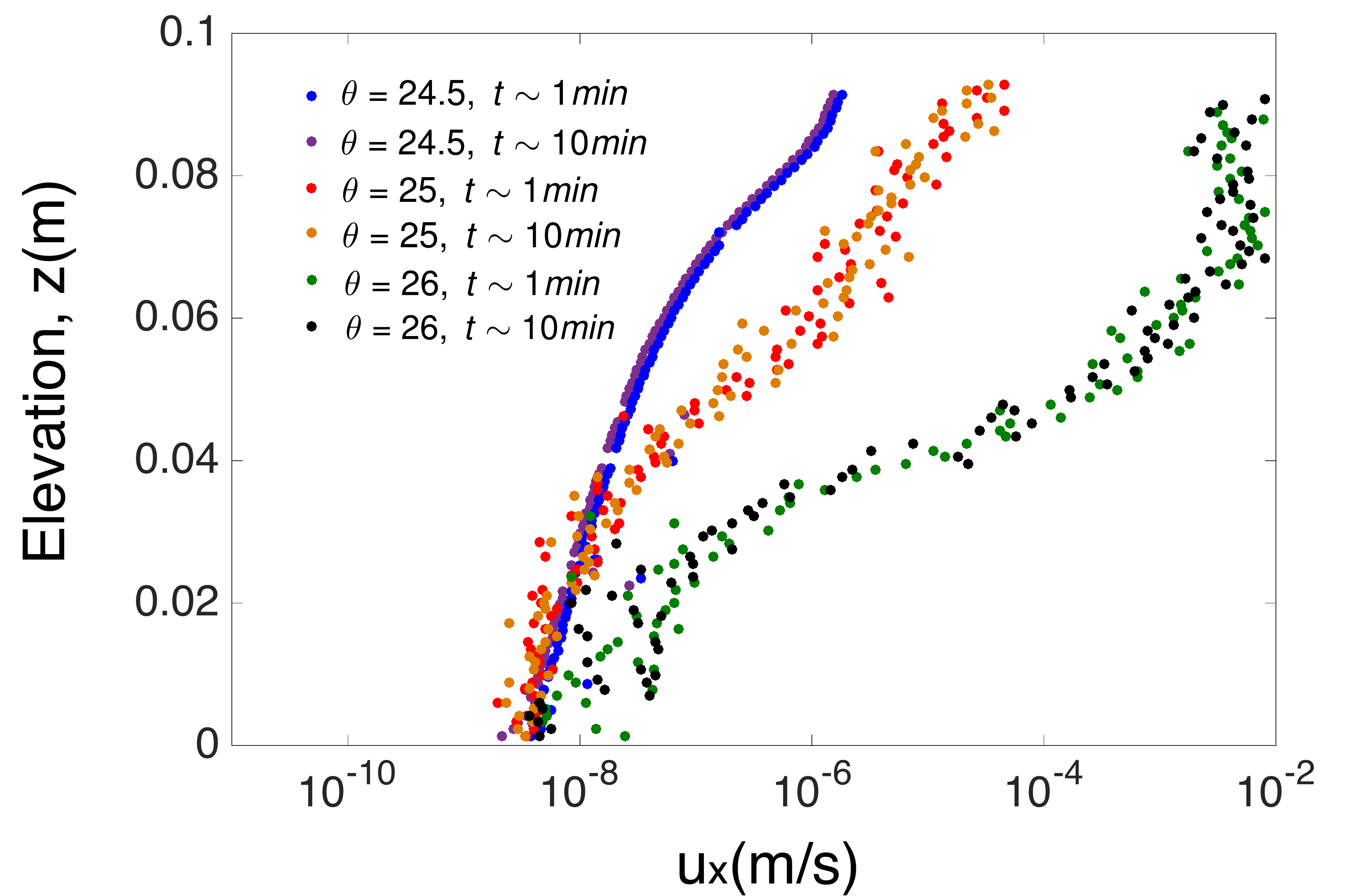}
\end{figure}
\noindent {\bf Fig. S1.} Time-averaged horizontal (x-dir) velocity ($u_x(z)$) profiles for the granular hillslope simulations at $\theta = [24.5, 25, 26] \degree$ for simulation run durations equal to 1 and 10 minutes. The velocity profiles show no significant change in rearrangement dynamics, and also that creep and slow motions persist over longer simulation times.

\clearpage

\begin{figure}
  \centering
      \includegraphics[width=1.0\textwidth]{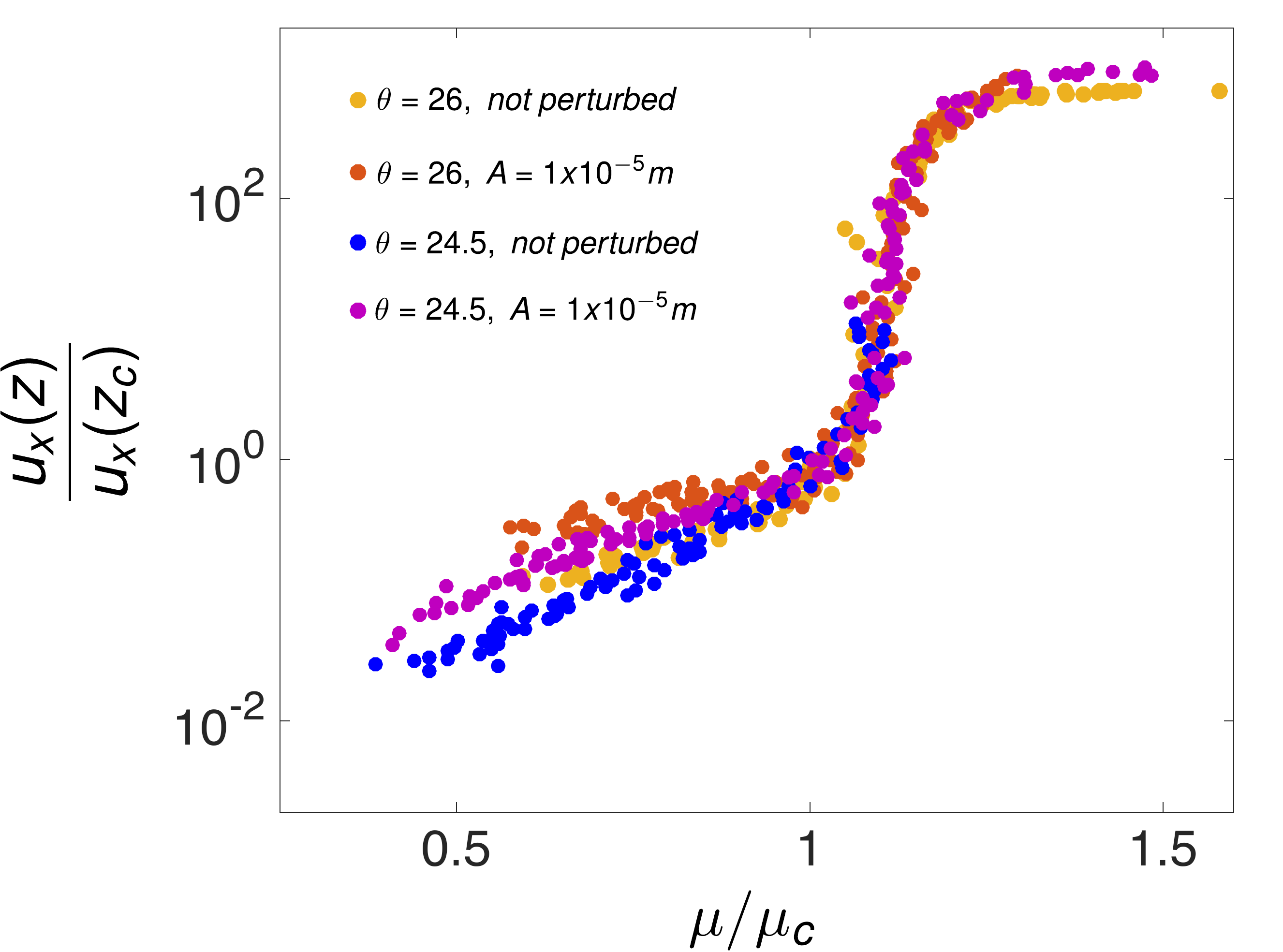}
\end{figure}
\noindent {\bf Fig. S2.} The variation of normalized local horizontal velocity ($\frac{u_x(z)}{u_x(z_c)}$) as a function of normalized local friction coefficient ($\frac{\mu}{\mu_c}$) in non-perturbed and perturbed simulations, at two different inclinations below and above the bulk angle of repose. When external perturbations are on, they are applied to grains in all directions and as a randomly distributed displacement within [-A,A] amplitude range. A maximum disturbance amplitude of $A = 0.005 \times$average grain diameter [$A = 1\times 10^{-5} m$] is used.

\clearpage

\begin{figure}
  \centering
      \includegraphics[width=1.0\textwidth]{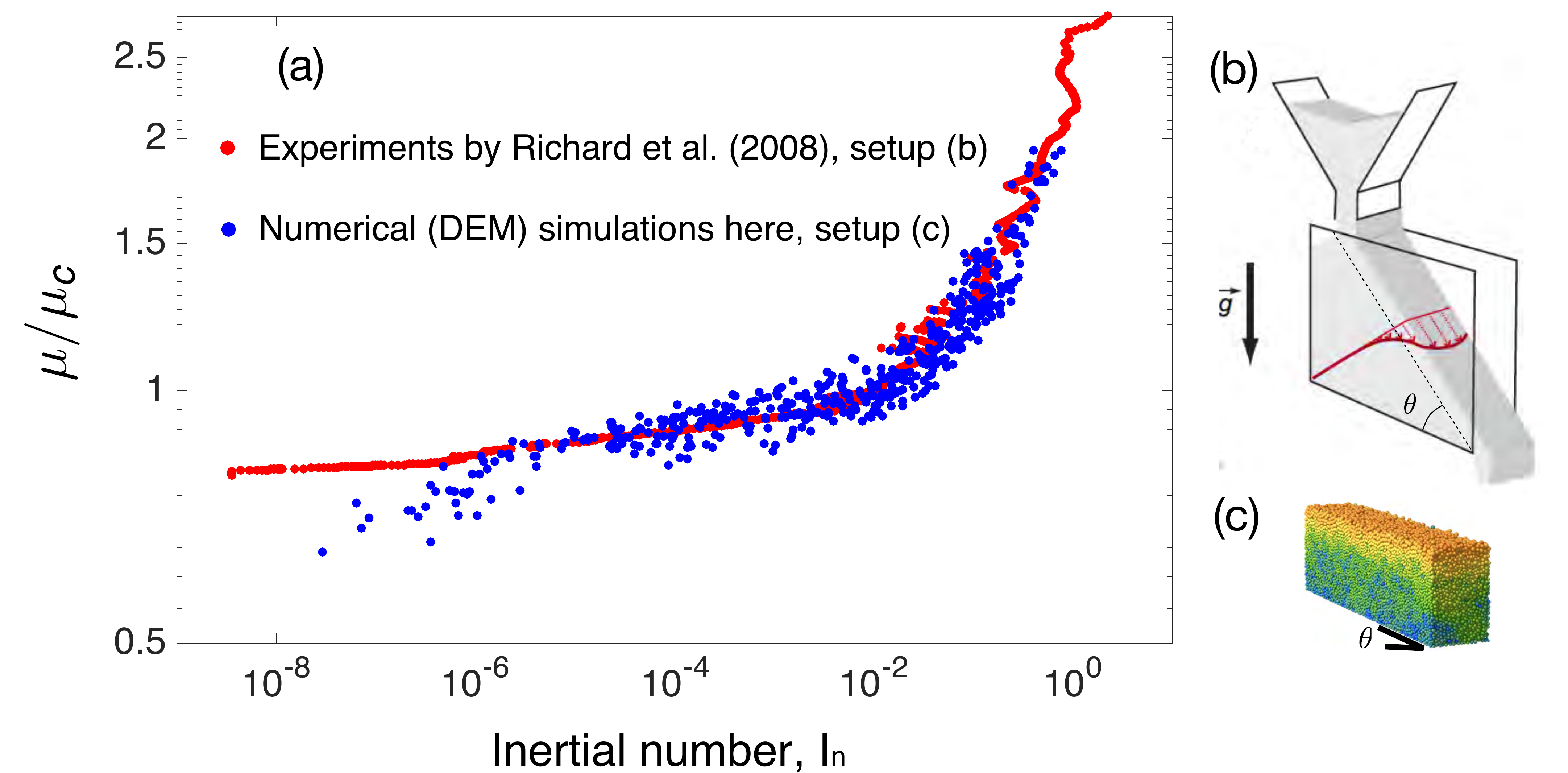}
\end{figure}
\noindent {\bf Fig. S3.} (a) The variation of normalized local friction coefficient, $\frac{\mu}{\mu_c}$, versus granular inertial number, $I_n = \frac{\dot\gamma \, d_{mean}}{\sqrt{\sigma_n/ \rho}}$, where $\dot\gamma$ is the granular shear rate, $d_{mean}$ is the average grain diameter, $\sigma_n$ is confining (normal) pressure, and $ \rho$ is grain density. Shown are results from a heap flow experiment reported by Richard et al. [ref. \cite{richard2008rheology}, setup shown in (b)] and our numerical simulations (setup in panel (c)). Results show the DEM simulation captures experimental rheology well, except for very low inertial numbers where it is known that creep deviates from the local rheology picture. 

\clearpage

\begin{figure}
  \centering
      \includegraphics[width=1.0\textwidth]{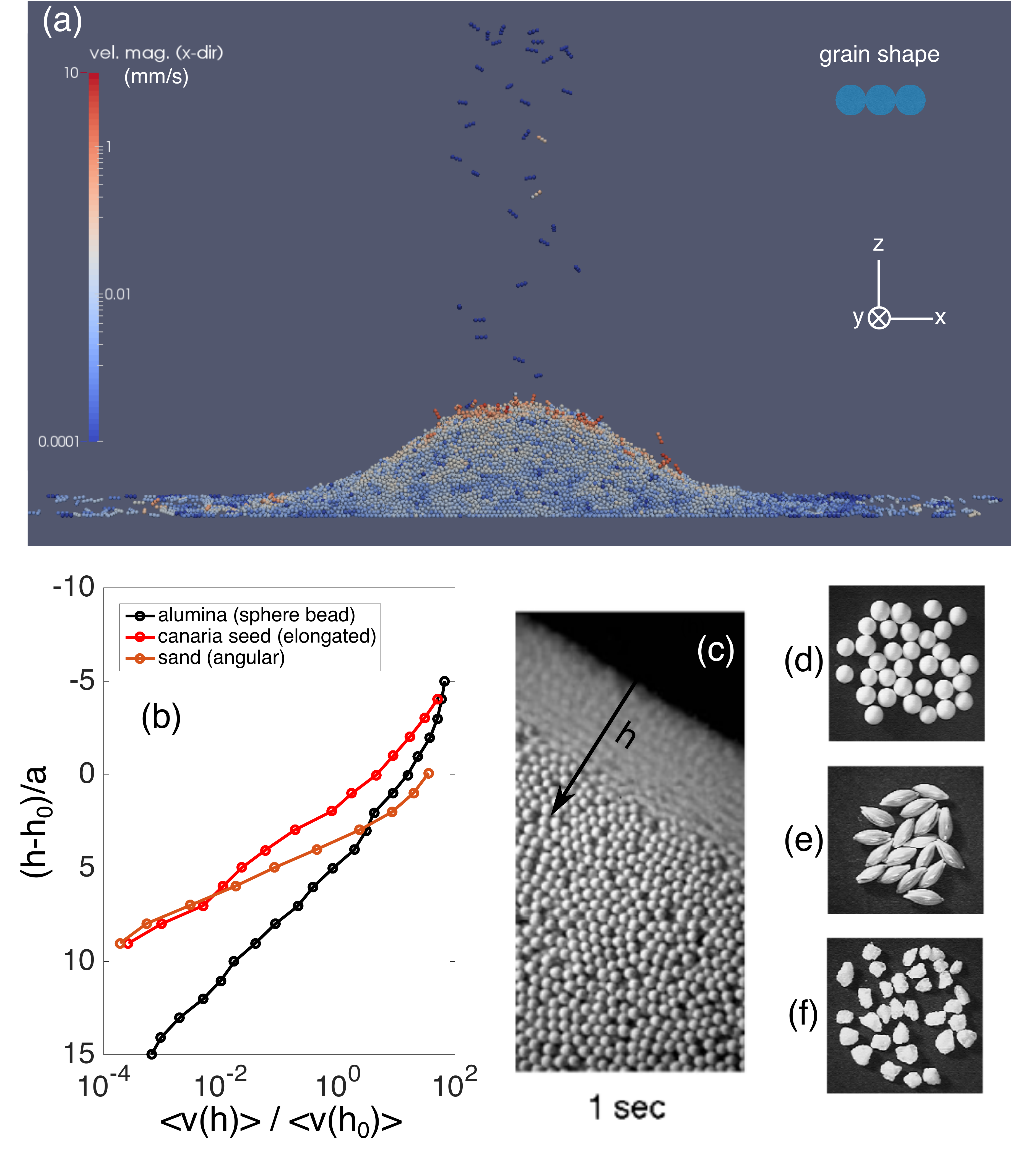}
\end{figure}
\noindent {\bf Fig. S4.} Demonstration that creep occurs for grains of various shapes. (a) Visualization of a three dimensional granular heap DEM simulation with elongated grains (confined between two side-walls, separated by $\sim$ 10 elongated grain size). The grains are colored according to the magnitude of horizontal (x-dir) instantaneous velocity. The color scale is logarithmic. (b) Mean horizontal velocity profiles in heap flow experiments reported by Komatsu et al. [ref. \cite{komatsu2001creep}] in a setup similar to Fig. S3-(b), with alumina spherical grains (d), canaria seed (e) and angular sand (f). In panel (b), $h$ is the depth measured from the surface of the granular layer, $h_0$ is the depth at which the transition from creep to dense flow takes place in each experiment, and $a$ is grain diameter. Panels (b)-(f) are reproduced from Komatsu et al. (2001)\cite{komatsu2001creep}.  

\clearpage

\begin{figure}
  \centering
      \includegraphics[width=1.0\textwidth]{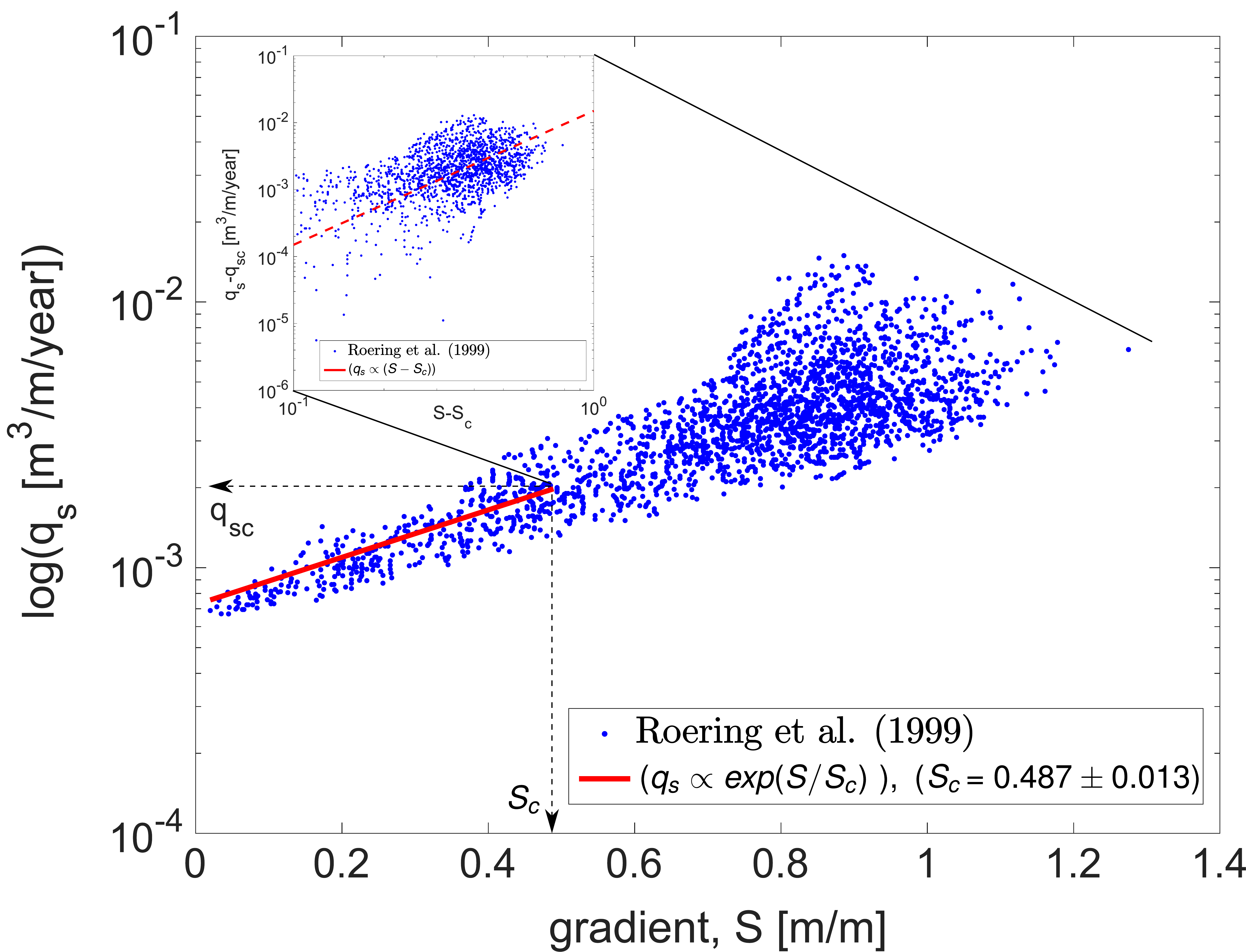}
\end{figure}
\noindent {\bf Fig. S5.}  Variation of volumetric sediment flux versus hillslope gradient in the study by Roering et al. (1999) \cite{roering1999evidence}. The inset shows the variation of sediment flux in excess of the sediment flux at the transition point ($q_s - q_{sc}$) versus hillslope gradient in excess of the critical gradient ($S-S_c$) in log-log scale. 

\clearpage

\begin{figure}
  \centering
      \includegraphics[width=1.0\textwidth]{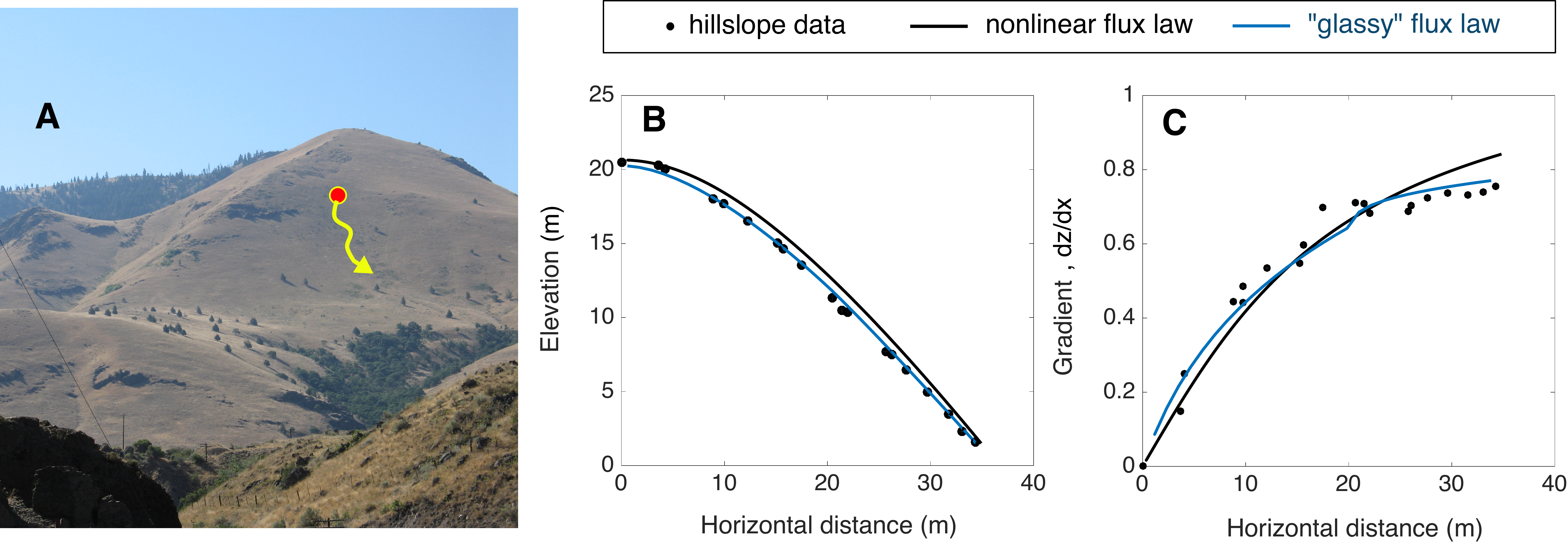}
\end{figure}
\noindent {\bf Fig. S6.}  (A) Example profiles of hillslopes in Deschutes river valley, Oregon (USA), and a schematic soil grain trajectory on the surface. Photo courtesy of Stephen Taylor, Western Oregon University. (B) and (C) show elevation-distance (surface profile) and gradient-distance profiles, respectively, of a hillslope measured in OCR \cite{roering2001hillslope}. The black data points in panels (B) and (C) are field measurements; black line shows prediction from the widely used non-linear hillslope diffusion equation \cite{roering1999evidence}, and blue line is our proposed ``glassy" flux model for the same hillslope from measured and estimated parameters presented by Roering et al. (2001)\cite{roering2001hillslope}. The glassy flux model used in this figure has the form proposed in Eq. (2), with $S_c = 0.5$, $\beta = 5/2$, and $A = 222$. For this small hillslope, both models perform well, although the glassy law is marginally better and predicts a flattening out of the hillslope at gradients larger than critical. For longer hillslopes as shown in Figure S8, the nonlinear diffusion law deviates substantially more.

\clearpage

\begin{figure}
  \centering
      \includegraphics[width=1.0\textwidth]{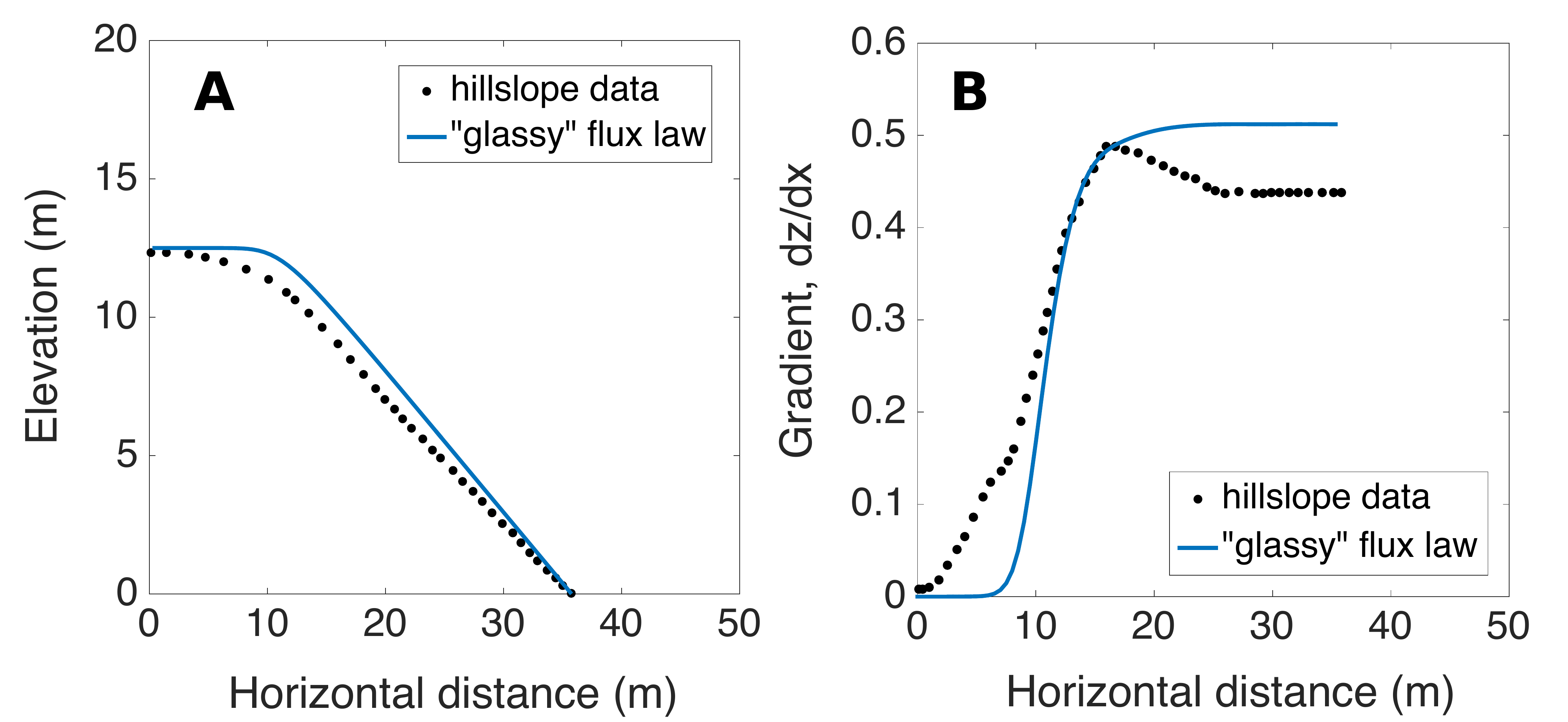}
\end{figure}
\noindent {\bf Fig. S7.} (B) Elevation-distance and (C) gradient-distance profiles, of a hillslope measured in Santa Ynez Valley in the tectonically active Transverse Ranges near Santa Barbara, California \cite{gabet2000gopher}. The black data points in panels (B) and (C) are field measurements, and blue lines in each panel demonstrate results of our proposed glassy flux model for the same hillslope from measured and estimated parameters presented by Gabet (2003) and (2000)\cite{gabet2003sediment,gabet2000gopher}. The glassy flux model used in this figure has the form proposed in Eq. (2), with $S_c = 0.513$, $\beta = 3/2$, and $A = 9010$. Here, we model the transient evolution of hillslope topography by iteratively solving equations (1) and (2) starting from a flat initial condition with constant uplift rate $C_o = 0.5 \, \text{mm/yr}$ for this region \cite{rockwell1984chronology}. The profile is evolved for 25,000 years, roughly the time needed to reach present relief \cite{rockwell1984chronology}. Additionally, about 16500 years BP, the Santa Barbara region emerged from the latest glacial maximum \cite{kennett199520}, rainfall diminished significantly,  and channel incision in the region no
longer had the capacity to remove sediment brought down from the hillslopes and keep up with the uplift rate. We therefore stop uplift/incision and run the model for another 16500 years following the first 25000 years uplift/incision simulation. These simulation steps and configurations are identical to the conditions implemented by Gabet (2000) \cite{gabet2000gopher} for predictions of the same hillslope profiles  in this region using linear (that includes gopher bioturbation effects) and nonlinear flux laws. Even with this complex geologic history, the glassy flux model reproduces the overall shape and relief of the hillslope reasonably well.

\clearpage

\begin{figure}
  \centering
      \includegraphics[width=1.0\textwidth]{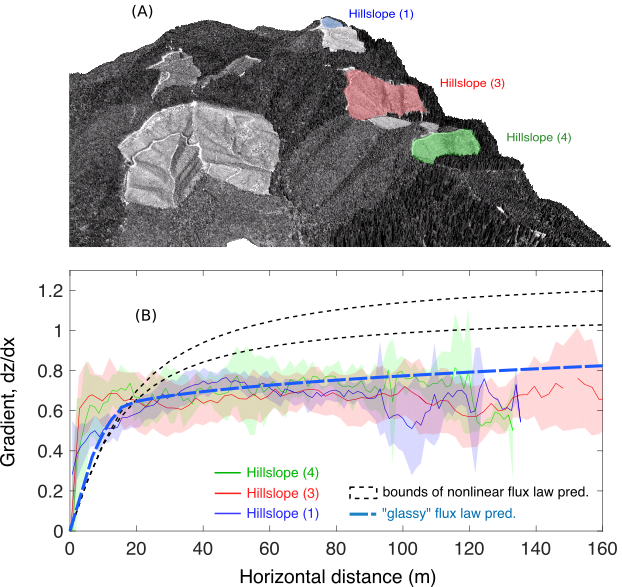}
\end{figure}
\noindent {\bf Fig. S8.} (A) Topographic map of a field area in the Oregon Coast Range, from airborne lidar data publicly available at ref.\cite{notemapfig5}. Shaded areas in green, red and blue colors show three different hillslopes examined here. (B) The gradient-distance profiles of hillslopes A-C, color-shaded in panel (A). The shaded bar for each measurement shows errorbar (one standard deviation). The black dashed lines show the bounds of the predictions using the nonlinear flux law \cite{roering1999evidence}, with range of diffusion coefficients and critical slopes estimated for OCR \cite{roering2007functional}. The blue dashed line is the prediction of glassy flux model with $S_c = 0.5$, $\beta = 5/2$. For these longer hillslopes (compared to Fig. S6), we see the glassy flux model is a better match to the data while the nonlinear law overpredicts hillslope gradients.

%


\begin{thebibliography}{86}
\bibitem [{not()}]{notemapfig5}%
  \BibitemOpen
  \href@noop {} {}\bibinfo {note} {The lidar measurements are carried out by
  NCALM Seed. The data source is available via opentopoID: OTLAS.052014.26910.3
  on opentopography platform.}\BibitemShut {Stop}%
\end{thebibliography}

\end{document}